\documentclass{IEEEcsmag}

\usepackage[colorlinks,urlcolor=blue,linkcolor=blue,citecolor=blue]{hyperref}
\expandafter\def\expandafter\UrlBreaks\expandafter{\UrlBreaks\do\/\do\*\do\-\do\~\do\'\do\"\do\-}
\usepackage{listings}
\usepackage{xcolor}
\lstdefinestyle{arxivstyle}{
    backgroundcolor=\color{white},
    basicstyle=\footnotesize\ttfamily,  % Small, typewriter font
    breaklines=true,                  % Wrap long lines automatically
    captionpos=b,                     % Position caption at the bottom
    commentstyle=\color{gray},        % Color for comments
    keywordstyle=\color{blue},        % Color for keywords
    stringstyle=\color{red},          % Color for strings
    numbers=left,                     % Line numbers on the left
    numberstyle=\tiny\color{gray},    % Style of line numbers
    rulecolor=\color{black},          % Frame border color (none if undefined)
    showspaces=false,
    showstringspaces=false,
    showtabs=false,
    stepnumber=1,                     % Line number step
    tabsize=2,
    frame=lines                       % Draw top and bottom lines only
}

\jvol{XX}
\jnum{XX}
\paper{8}
\jmonth{Feb}
\jname{Publication Name}
\jtitle{Publication Title}
\pubyear{2026}

\begin{document}

\sptitle{Article Type: Feature/Perspective}

\title{Fog Computing and Large Language Models: A vision for mutual beneficiaries}

\author{Satish Narayana Srirama}
\affil{School of Computer and Information Sciences, University of Hyderabad, India}

\markboth{FEATURE}{Fog Computing and LLM Conglomeration}

\begin{abstract}\looseness-1Fog computing utilizes proximal computational resources for sensor data processing and actuation, and addresses the latency, network load, and privacy issues of cloud-centric Internet of Things. On the other hand, Large Language Models (LLMs) are a type of deep learning AI models, which are trained on enormous text data, that perform various natural language processing tasks such as translation, question answering, text summarization, and code generation. LLMs are generally cloud-centric, requiring abundant GPU memory and computing capabilities, again face the same issues that led to fog computing. This pushes the necessity for LLM support in the proximity on fog infrastructure, requiring LLM optimizations such as parameter-weight quantization, pruning, low-rank adaptation etc. Meanwhile, fog computing also gets benefit from LLM’s ability for code generation, in the dynamic deployment of fog-based applications. The paper addresses how both fog computing and LLMs can be mutual beneficiaries, discussing the state-of-the-art and future research scope.
\end{abstract}

\maketitle
%\copyrightnotice

\section{Introduction}\label{sec:intro}

Recent advances in the Internet of Things (IoT) and the growing demand for real-time applications across domains such as smart homes, cities, transportation, and healthcare have driven an unprecedented proliferation of connected devices, ranging from embedded sensors and smart gadgets to smartphones and industrial IoT platforms. These sensing-enabled physical objects, connected over the Internet, collectively form the foundation for realizing increasingly sophisticated smart applications. For realizing these smart applications, the data collected from the connected devices is generally pushed to the cloud for processing and the control signals are sent back to the devices for the required actuation. To address the latency, network load, and privacy issues of such cloud-centric IoT applications, fog computing was coined, which utilizes proximal computational resources for sensor data processing and actuation. Since its proposal, fog computing has attracted significant attention and the research community has focused on addressing different challenges~\cite{srirama2024decade} such as fog frameworks, simulators, resource management, placement strategies, quality of service (QoS) aspects, mobility, edge analytics, fog economics etc. 

Meanwhile, Large Language Models (LLMs)~\cite{vaswani2017attention}, a type of deep learning artificial intelligence (AI) models, which are trained on vast amounts of text data to understand and generate human language, have evolved. LLMs can perform various natural language processing (NLP) tasks such as translation, question answering, text summarization, and code generation. Building on these advancements, multimodal LLMs have emerged, integrating text with additional modalities such as images, audio, and video to support richer and more context-aware inference. By unifying multiple information streams within a single model, these multimodal variants extend LLM utility beyond text-centric applications and enable more comprehensive real-world interactions, as part of IoT applications. 

LLM-supported IoT applications are rapidly expanding across several domains, driven by the growing integration of intelligent reasoning capabilities within connected environments~\cite{RAY2025275}. In smart homes, LLMs enable context-aware assistants capable of interpreting user intent and autonomously controlling appliances, managing their usage and thus saving energy and improving sustainability, and enhancing home security through analysis of sensor logs and camera feeds. Smart city deployments can leverage LLMs for real-time citizen interaction, automated incident summarization from heterogeneous sensor streams, and the generation of situational insights for urban operations. In industrial IoT settings, LLMs provide natural-language diagnostics derived from machine sensor data, support workflow and procedure generation, and automate compliance reporting. Healthcare IoT systems benefit from LLM-generated clinical summaries~\cite{van2024adapted} of wearable and ambient sensor data, personalized adherence guidance and multimodal triage support that combines diverse data sources such as physiological data (such as heart rate, respiratory rate, oxygen saturation, and body temperature) and medical images (such as chest X-rays, CT scans etc.) to provide a comprehensive assessment~\cite{shukla2026benchmarking, wang2026evaluation}. 

LLMs in general are again cloud-centric, requiring abundant GPU (Graphics Processing Unit) memory and computing capabilities, thus face the same issues such as the latency, network load, and privacy, that led to the fog computing. This pushes the necessity for LLM support in proximity on fog infrastructure. Such on-device and near-edge LLMs enable privacy-preserving analytics, local summarization, and natural-language coordination among distributed IoT nodes, thereby strengthening autonomy and real-time responsiveness across IoT ecosystems.

However, fog devices have strict memory and computation constraints, making standard Transformers used in LLMs, such as GPT-4 or Llama 70B that typically rely on cloud-scale GPU or TPU (Tensor Processing Unit) infrastructure and high-precision inference, too heavy for such fog devices. Consequently, a range of approaches has emerged to enable the deployment of LLMs, typically Small Language Models (SLMs), which are 1B-7B parameter range, within such resource-limited environments. Model compression techniques such as quantization, pruning, and low-rank adaptation, reduce parameter size and computation overhead while preserving functional accuracy. Lightweight architectural variants, including linear-attention and recurrent memory augmented Transformers, further reduce computational and memory overhead associated with quadratic attention, making inference feasible on devices with limited RAM and bandwidth~\cite{xu2024device}. 

Meanwhile, fog computing also benefits from LLMs due to LLM’s ability for code generation. While fog computing is shown to be promising in wide range of applications, adoption rate of fog computing is not yet in proportion with the advantages it proposes~\cite{srirama2024decade}. Some of the main challenges are due to the resource-constrained nature and heterogeneity of fog devices, and lack of standardization in building and deploying fog applications. Currently, fog services are deployed on the devices manually, using application-specific proprietary solutions. There is significant push to adapt standardization efforts such as OASIS Topology and Orchestration Specification for Cloud Applications (TOSCA), thereby supporting standards-based dynamic deployment and management of fog applications. For this type of dynamic deployment, the templates for defining the fog application configuration are generated manually, by following the DevOps model of Infrastructure as Code (IaC), and the applications are deployed using the orchestrators. With the LLMs, the research community is attempting to automatically generate these templates, thereby supporting intelligent dynamic deployment of fog-based applications. 

The paper addresses both the aspects of how fog computing and LLMs can be mutual beneficiaries of each other, and provides a vision for the future research scope. % The rest of the paper is organized as follows. %discussing the state-of-the-art, demonstrating feasibility and some performance analytics,  
%Section 2 discusses the optimizations required for achieving on-device LLMs. Section 3 discusses the fog application template generation with LLMs for dynamic deployment. Section 4 provides further research scope for the fog and LLM conglomeration and Section 5 provides a conclusion for the paper.  
The contributions of the paper are:
\begin{itemize}
\item Provides a detailed discussion on the optimizations required for achieving on-device LLMs, along with experimental results showing impact of quantization on LLM models on Raspberry PI devices. 
\item Discusses the need and directions for fog application template generation with LLMs, along with a case study that exploits the IaC service templates for dynamic deployment. 
\item Provides further research scope for the fog and LLM conglomeration.
\end{itemize}

\section{Optimizations and feasibility of LLMs for fog devices}
\subsection{LLM}

At the foundation of contemporary LLMs lies the transformer architecture, which processes input in the form of tokens. A token represents a basic linguistic unit, such as a word, and is converted into a vector embedding through tokenization, thereby homogenizing the heterogeneous textual input into a computation suitable format. The transformer architecture then employs several fundamental mechanisms to learn dependencies across the input sequence. The first, self-attention, enables each token to evaluate its relevance with respect to all other tokens that have appeared before, capturing contextual and semantic relationships. The second, multi-head attention, extends this capability by executing several self-attention operations in parallel, with each head modelling different aspects of the input. Beyond attention, each transformer layer also includes a multilayer perceptron (MLP) block, typically consisting of two linear transformations with a non-linear activation. This MLP component further refines token embeddings by enabling complex feature transformations that are difficult to capture through attention alone. The transformer architecture comprises multiple stacked layers of attention and MLP blocks, e.g. up to 96 in the case of GPT-3 model variants, allowing the network to progressively refine contextual representations and capture increasingly abstract linguistic patterns.  Together, these mechanisms allow the model to learn both short- and long-range dependencies, supporting the generation of coherent and contextually rich text. 

Structurally, transformers comprise an encoder–decoder architecture, although several modern LLMs adopt decoder-only variants for autoregressive generation. In the classical architecture, the encoder processes the input sequence, derives contextualized representations, and encodes relationships among tokens using self-attention. These contextual embeddings are then used by the decoder, which generates the output sequence incrementally. The decoder incorporates encoder–decoder attention to identify and prioritize the most relevant components of the encoded input for each token it generates. For every token, the model computes Query, Key, and Value vectors: the Query expresses the information being sought, the Key characterizes the token’s attributes, and the Value supplies the associated content. The interaction between Query and Key vectors yields attention scores that quantify the influence of one token on another, while the Value vectors propagate the relevant information forward through the network~\cite{vaswani2017attention}.

These computations are repeated across multiple layers, with the resulting Key and Value tensors stored as part of the key–value (KV) cache. The KV cache accelerates inference by allowing previously computed contextual information to be reused for subsequent tokens, reducing redundant computation during generation. However, this benefit incurs substantial memory overhead, as the cache grows with the number of layers and the length of the input sequence; both the memory footprint and computational requirements of attention scale quadratically with sequence length. In practice, the KV cache for a single request may reach gigabyte-level sizes. Without effective management strategies, this can quickly exhaust available memory, limiting the number of concurrent inference requests and degrading overall responsiveness. This problem is elevated further in resource constrained environments such as edge and fog computing. 

\subsection{Multimodal LLM}

With the advancements in LLMs, multimodal LLMs have emerged, which can process and reason with multiple modalities, i.e. types of data, such as text, images and audio~\cite{Varughese:MLLM2026}. Each of these modalities has its own structure and requires divergent ways to represent and interpret information. Text can be represented as a sequence of words, an image as a grid of pixels and audio as a continuous waveform or spectrogram. These representations are then transformed to the dense vector representations that capture semantic information, such as the tokenization discussed for textual data. With images, advanced architectures such as vision transformers (ViT) or convolutional neural networks (CNNs) are employed to extract visual features such as shapes, colours and spatial patterns. With audio, specialized encoders such as wav2vec or HuBERT~\cite{hsu2021hubert} are employed that process raw waveforms to produce representations of speech or sound cues. These abstract features are then mapped into a shared embedding space i.e. again into vectors; thus, the model combines the vectors to form a unified multimodal representation. Once the merged features are processed, the model produces an output that answers a specific task handled by an output decoder. For example, the decoder can generate image captioning or video description.

\subsection{On-device LLMs}

Building upon the architectural principles described above, recent research has increasingly focused on enabling LLMs to operate directly on resource-constrained edge and mobile devices. This shift toward on-device LLMs stems from the need to reduce latency, enhance data privacy, and improve service availability, the factors that closely parallel the motivations behind fog and edge computing. On-device LLMs rely on a combination of model compression techniques, architectural optimizations, and hardware-software co-design. Techniques such as post-training quantization (INT8, INT4, and even sub-4-bit formats), structured and unstructured pruning, knowledge distillation, and low-rank adaptation are essential in shrinking model footprints while preserving the task performance. These methods reduce parameter counts, memory bandwidth requirements, and arithmetic complexity, allowing models with hundreds of millions of parameters to execute in real-time on devices with only a few gigabytes of RAM. 

\begin{figure*}
\centering
\includegraphics[width=0.95\textwidth]{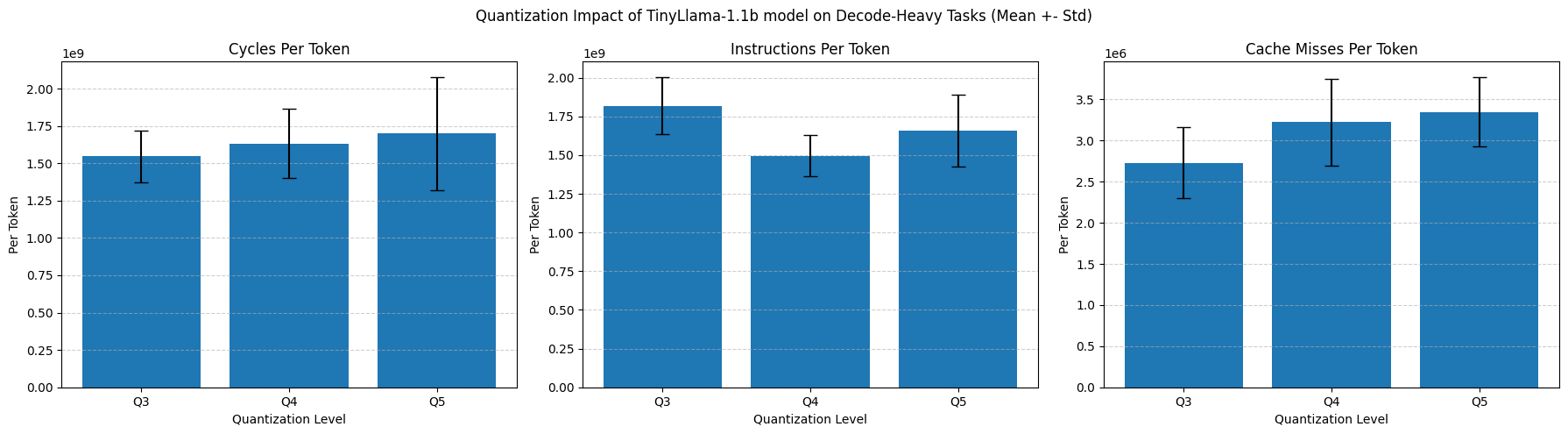}
\caption{Quantization impact on TinyLlama-1.1b model for decode-heavy tasks}
\label{fig_QuantizationImpact}
\end{figure*}

For example, Figure~\ref{fig_QuantizationImpact} shows the impact of quantization on TinyLlama-1.1b model for decode-heavy tasks (tasks with less input tokens but asked to generate high number of output tokens) on a Raspberry PI 4 Model B with 8 GB SDRAM, with active cooling to remove the effect of thermal throttling. The experiments were conducted on Raspberry Pi OS 64-bit using llama.cpp with pre-quantized GGUF models obtained from Hugging Face, while Linux \texttt{perf stat} was used for performance measurements. Here, Q3, Q4, and Q5 refer to post-training quantization levels where the number denotes the approximate bits-per-weight used to represent the model weights, i.e. 3-bit, 4-bit, and 5-bit, respectively. Higher quantization levels preserve more of the original FP16 precision. The quantized variants Q3\_K\_L (3-bit, large K-quant), Q4\_K\_M (4-bit, medium K-quant), and Q5\_K\_M (5-bit, medium K-quant) are used for the experiments. The left, center, and right plots in the figure represent cycles per token, instructions per token, and cache misses per token, respectively, where lower values indicate improved efficiency. Across all three metrics, Q3 consistently performs best, suggesting that lower-bit quantization can substantially reduce memory-access pressure and computational overhead for decode-heavy workloads on fog-class devices. For extracting the results of cycles, instructions and cache misses per token, 9 different tasks of the same regime (decode-heavy tasks) were used, and the reported results correspond to averages across all these tasks. There are no FP16 or FP32 baselines in the figure, as the comparison is limited to different GGUF quantization levels. Without quantization, the model cannot practically run on the fog device. For the experiments shown in Figure~\ref{fig_QuantizationImpact}, Q3, Q4, and Q5 achieved mean throughputs of 3.45 ± 1.26, 3.20 ± 1.05, and 2.76 ± 0.98 tokens per second, respectively. Thermal behavior was monitored using \texttt{vcgencmd}, with all quantization levels operating at a mean temperature of approximately 58-59°C and a maximum observed temperature of 63.7°C, without any CPU throttling or under-voltage events. Although direct energy measurements were not available due to the absence of external power instrumentation, the results indicate stable execution under sustained workloads.

In addition, techniques such as operator fusion, optimized kernels, and hardware-aware compilation pipelines (e.g., through TensorRT, TVM, or custom NPU runtimes) significantly accelerate model execution on heterogeneous edge hardware. Offloading strategies and collaborative inference across edge–fog–cloud continuum layers also help distribute computational load, maintaining low latency while respecting energy and memory budgets~\cite{xu2024device}. Collectively, these optimizations can make on-device LLMs increasingly practical for fog environments, enabling intelligent, privacy-preserving analytics directly at the network edge. However, several of these optimizations at the moment are targeted at reducing the latency and resource usage of LLMs in general and their applicability on the resource constrained fog devices is to be studied and understood in detail. Further, these optimizations should be tailored and trained for the specific requirements of the fog-based applications.

%\section{Optimizations and feasibility of LLMs for fog devices}

%To study the effects of the LLM optimizations on general-purpose edge/fog devices such as the Raspberry PI, we conducted the following experiments. 
%Have graphs with few optimizations of LLMs on Raspberry Pis

%All these optimizations have profound impact on..

%These optimizations should be further tailored and trained for the specific requirements of the fog-based applications. 

\section{Dynamic fog application template generation with LLMs}

Some of the main challenges associated with fog adoption in different real-time applications lies with the heterogeneity, interoperability and lack of standardization issues~\cite{srirama2024decade}. Fog devices include microcontrollers such as Raspberry Pis, drones, gateway devices, network switches, routers etc. Each of these devices will have their own hardware and software-support configurations. An application/service developed for a particular fog device cannot be easily ported to another device. These portability challenges can be addressed, provided, the applications are developed for a standard/widely-adopted platforms such as docker containers, which can therefore be ported to compatible/supporting devices. The fog nodes then have to interact among themselves for offloading or collaborative execution of tasks, leading to interoperability challenges. These interoperability challenges can be addressed with container orchestration solutions such as Docker swarm or Kubernetes. Still the applications are to be designed and developed for the targeted devices and have to be manually configured either locally or through remote connections, which is not easy for large-scale and geographically distributed IoT application deployments.

To support automated configuration and dynamic deployment of fog computing applications, specifications such as OASIS TOSCA can be adapted, following the IaC methodology. TOSCA follows the model-driven approach and provides a language for defining a service topology that describes application components and their relationships, and for specifying the lifecycle management actions such as creation, modification, and deletion, of the services using orchestration processes~\cite{Lauwers:TOSCA2.0}. FogDEFT~\cite{basak2024fog} framework extended the TOSCA specification for fog computing applications, so that applications can be dynamically deployed on the specified infrastructure through orchestrators such as xOpera. The fog application blueprint is to be defined as the TOSCA service template, describing application topology and components, while the associated Ansible scripts are used by the orchestrators to configure, install software and manage the lifecycle. 

\begin{figure}
\centering
\includegraphics[width=0.49\textwidth]{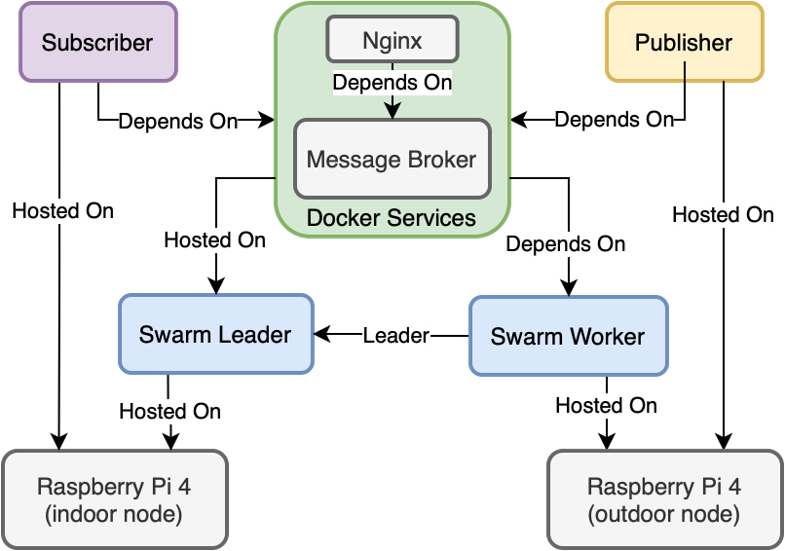}
\caption{Climate control system - Design blueprint}
\label{fig:fogapplication}
\end{figure}

\begin{lstlisting}[caption={Climate control system - TOSCA service template}, label={lst:yaml}]

    ---
    tosca_definitions_version: tosca_simple_yaml_1_3
    
    description: Service Template for climate control
    
    metadata:
      template_name: "ClimateControl Service Template"
      template_author: "Cloud & Smart Lab"
      template_version: "2.0"
    
    imports:
      - relationshiptypes/token_transfer/token_transfer.yaml
      - nodetypes/swarm_leader/swarm_leader.yaml
      - nodetypes/swarm_worker/swarm_worker.yaml
      - nodetypes/docker_services/docker_services.yaml
      - nodetypes/docker_containers/docker_containers.yaml
    
    topology_template:
      inputs:
        outdoor:
          type: string
          description: Input IP address of outdoor node
        
        indoor:
          type: string
          description: Input IP address of indoor node
    
      node_templates:
        # All fog nodes
        outdoor-node:
          type: tosca.nodes.Compute
          attributes:
            private_address: { get_input: outdoor }
            public_address: { get_input: outdoor }
    
        indoor-node:
          type: tosca.nodes.Compute
          attributes:
            private_address: { get_input: indoor }
            public_address: { get_input: indoor }
    
        # Swarm leader node
        docker-swarm-leader:
          type: fog.docker.SwarmLeader
          requirements:
            - host: indoor-node
    
        # Swarm worker nodes
        docker-swarm-worker:
          type: fog.docker.SwarmWorker
          requirements:
            - host: outdoor-node
            - leader: docker-swarm-leader
    
        # Docker Service (SWARM)
        broker-service:
          type: fog.docker.Services
          properties:
            name: broker
            url: https://repo/broker/docker-compose.yaml
          requirements:
            - host: docker-swarm-leader
            - dependency: docker-swarm-worker
    
        # Standalone containers
        sensor-data-publisher:
          type: fog.docker.Containers
          properties:
            name: publisher
            url: https://repo/publish/docker-compose.yaml
          requirements:
            - host: outdoor-node
            - dependency: broker-service
    
        actuator-data-subscriber:
          type: fog.docker.Containers
          properties:
            name: subscriber
            url: https://repo/subscribe/docker-compose.yaml
          requirements:
            - host: indoor-node
            - dependency: broker-service
    
\end{lstlisting}

% \begin{figure}
% \centering
% \includegraphics[width=0.49\textwidth]{FogDEFTAppTemplate.png}
% \caption{Climate control system - TOSCA service template}
% \label{fig_QuantizationImpact}
% \end{figure}

% \begin{figure}[!t]
% \centering
% \begin{subfigure}%{.5\textwidth}
%     \centering
%     \includegraphics[width=0.37\columnwidth]{FogDEFTApp.png}
%     %\caption{ Average time taken (seconds) by the framework per round}
%     \label{fig:blueprint}
% \end{subfigure}
% \hspace*{0.25in}
% \begin{subfigure}%{.5\textwidth}
%     \centering
%     \includegraphics[width=0.57\columnwidth]{FogDEFTAppTemplate.png}
%     %\caption{Models accuracy on both synchronous and asynchronous updates }
%     \label{fig:template}
% \end{subfigure}

% \caption{Climate control system (a) Design blueprint (b) TOSCA service template}
% \label{fig:fogapplication}
% \end{figure}

Figure~\ref{fig:fogapplication} shows the service design blueprint of a fog application, climate control system of a convention centre. The application uses different sensors (temperature, humidity, light, barometric pressure, proximity, and microphone) to sense the environment and relatively performs actuation tasks such as opening the doors and windows, managing the air-conditioning system etc. to improve the ambience in the convention centre. The bottom two grey nodes of the figure represent the fog nodes (Raspberry Pi). The green node in the middle represents Docker Services (Message broker and web server to show the sensor data and state) running in the swarm mode (interservice dependencies are mentioned in the docker-compose file). This Docker Service node is hosted and depends on two blue nodes, Swarm Leader and Swarm Worker, respectively. The relationship between the leader and worker is also shown in the figure. The remaining two yellow and purple nodes are Publisher and Subscriber services, respectively. The Publisher node pushes the sensor data collected from the devices to the message broker. The Subscriber node receives the broadcast of each update from the message broker and makes the respective adjustments to the actuators. Listing~\ref{lst:yaml} shows the TOSCA service template of the application. Through the service template and the orchestrator, the application can be dynamically deployed on any fog infrastructure and migrated very easily. More details of the framework, application and deployment procedure are available at~\cite{basak2024fog}. 

While the TOSCA templates support the dynamic deployment of fog computing applications, the templates are to be manually generated for each application, which is a tedious process. In addition, to achieve true intelligent fog computing applications, based on some AI inferences, the current configurations can be modified dynamically and new services can be deployed on the fog nodes. Although the service template approach can support such intelligent dynamic deployment, manual generation of the template becomes the bottleneck. The ability of LLM to automatically generate code has been studied extensively in the literature and is widely used in the industry~\cite{joel2024survey}. 

To support the automatic generation of TOSCA service templates, we are currently studying the ability of LLM to automatically generate such templates. LLMs are being trained with the RADON Particles repository~\cite{RADON:particlesGit} that contains the TOSCA service templates for a wide range of cloud-based applications and architectures, which were generated as part of our EU H2020 RADON~\cite{casale2020radon} project. The generated IaC can be evaluated with code generation benchmarks such as IaC-Eval~\cite{kon2024iac}. In addition, the generated TOSCA service templates are to be validated with the syntactic and semantic validation features of the orchestrators, e.g. xOpera has such a validator. To improve the semantic intent of the generated code, techniques such as Retrieval-Augmented Generation (RAG) or agentic feedback loops can be employed to attain better accuracy~\cite{nekrasov2025iac}. 

In our preliminary studies of automatic service template generation, the validation process progresses through three iterative levels. Syntax validation is performed via the \texttt{"opera validate"} operation, where error messages (such as malformed interface blocks, missing capability bindings, and unresolved attribute references) are continuously included in the LLM prompt to create an iterative fix until the template is validated and correctly parsed at each iteration. Semantic validation is achieved through RADON Particles, which are registered at the point of generation and prevent any nodes from being generated that do not exist in the RADON repository. Safety of deployment is provided by a second LLM instance that reviews the validated template on a higher level of intent than "opera validate", by checking for MQTT-oriented ordering issues, persistent issues, and issues associated with ARM incompatibilities. As a result of these three methods and the involved human monitoring (in addition to putting the right prompts, the programmer also has to manually modify the environment-dependent configuration details such as substituting actual IP addresses and usernames into \texttt{inputs.yaml}), we can prevent hallucinated configurations, unsafe actuation, and incorrect deployment decisions. The procedure needs further exploration. In addtion, achieving true intelligent fog computing applications, that take care of automated generation, verification and dynamic deployment, needs significant future research, and probably in the directions of Agentic AI~\cite{donta2025socio}.

\section{Further research scope in fog and LLM conglomeration}

We already discussed the scope of LLM optimizations for fog computing and utilizing LLM for automatically generating fog application templates, thus both domains getting benefit from each other. In addition to these, there are several other research directions that need further exploration. 
\begin{enumerate}
    \item \textbf{Layered LLM inferencing in edge-fog-cloud continuum}: With edge-fog-cloud continuum, for realizing IoT applications, the sensor data collected at the edge layer would move across the continuum with intermediate processing across the layers, until it is finally stored and analysed at the cloud, and respective control signals are sent back to the edge devices. The IoT applications can use local LLM inference, using smaller or specialized models at the fog devices, and the cloud can perform the full-scale LLM. The cloud can also manage model updates at the fog devices, by periodically pushing the updated models back to fog and edge layers. Thus, the data and requests would move upward along the continuum for processing, while results, insights, and updated models propagate downward, having significant scope for further research. 
    \item \textbf{LLM task scheduling on fog topology}: The LLMs at the fog nodes can act as decision points for task offloading, determining whether the requests can be served locally or should be escalated to the cloud based on different QoS aspects such as latency constraints, workload, and resource availability. Fog placement strategies are studied extensively~\cite{srirama2024decade}, and the knowledge can be adapted to the LLMs in dealing with dynamic task routing to fog nodes. This would eliminate the need for another dedicated offloading scheduler on the fog devices, when the targeted system/application architecture already demands on-device LLMs. 
    \item \textbf{Applications specifically taking advantage of LLM}: While there exist pilot case studies, more research should focus on developing IoT applications that can exploit the abilities of LLM, in domains such as smart healthcare and industrial IoT. 
    \item \textbf{Incentive models for fog resource providers}: While LLMs on fog devices lead to interesting applications, to encourage the fog nodes to participate in such applications, ideal incentive mechanisms for the fog resource providers are to be studied and employed.
    \item \textbf{LLM training through FL}: Federated learning (FL) is a decentralized machine learning method where a shared model trains across multiple devices, with each device training on local data, and model updates are sent to a central server for aggregation, and the aggregated model is sent back to the devices for next round of training. FL supports preserving privacy and security by sending only model updates and not the raw data. Similar sort of training can be adapted in the LLM training on local devices and exchanging only trained model to the centralized nodes for aggregating global LLMs. In addition, considering the resource-constrained nature of the fog devices, the FL may not consider the full LLM layers but only the last few layers for training, suitable for cases such as LLM fine-tuning, adapter training, or personalization. This training the last few layers is a cost-effective adaptation strategy that freezes the core foundation weights and updates only the task-specific parameters, enabling the model to learn individual user preferences and conversational styles, while significantly reducing computation and data requirements. %This approach will address the privacy issues of LLMs, which may be ideal in scenarios such as personalized medication. 

\end{enumerate}

\section{CONCLUSION}\label{sec:conclusion}
Fog computing and LLM conglomeration has the potential to address the latency, network load, and privacy issues of respective LLM-integrated IoT applications. Both fog computing and LLMs can be mutual beneficiaries, with fog computing supporting the on-device LLMs through various optimizations and LLM helping in dynamic template generation for truly intelligent fog computing applications. The conglomeration has further scope for research in layered LLM inferencing in edge-fog-cloud continuum and in designing real-world LLM-integrated IoT applications.

\section{ACKNOWLEDGMENTS}
 We thank financial support to UoH-IoE by MHRD, India (F11/9/2019-U3(A)). The author also thanks his students, R. Siyanwal, S. Basak, and S. Muthyala, who have contributed in producing the respective frameworks and results, over the years.

\def\refname{REFERENCES}

%\bibliographystyle{IEEEtran}
%\bibliography{references}
% Generated by IEEEtran.bst, version: 1.14 (2015/08/26)

\begin{IEEEbiography}{Satish Narayana Srirama}{\,} is a Professor at the School of Computer and Information Sciences, University of Hyderabad, India. He is also a Visiting Professor and the honorary head of the Mobile \& Cloud Lab at the Institute of Computer Science, University of Tartu, Estonia, which he led as a Research Professor until June 2020. His current research focuses on cloud computing, mobile web services, mobile cloud, Internet of Things, fog computing, migrating scientific computing and enterprise applications to the cloud, and large-scale data analytics on the cloud. He is an IEEE Senior Member, and an Editor of Wiley Software: Practice and Experience, a 56-year-old Journal. Contact him at satish.srirama@uohyd.ac.in.\vspace*{8pt}
\end{IEEEbiography}

\end{document}